\documentclass[aps,prd,twocolumn,nofootinbib]{revtex4}
\usepackage{graphicx,bm}
\usepackage{amsmath,amssymb}

\newcommand{\f}{{h}}
\DeclareMathOperator{\im}{Im}
\DeclareMathOperator{\re}{Re}

\newcommand\pd{\partial}
\newcommand\z{\zeta}

\begin{document}

\title{Towards the Gravity Dual of Quarkonium in the Strongly Coupled QCD Plasma}

\author{Hovhannes~R.~Grigoryan}
\affiliation{Physics Division, Argonne National Laboratory, Argonne, IL 60439-4843, USA}
\author{Paul~M.~Hohler}
\affiliation{Department of Physics, University of Illinois, Chicago,  IL 60607-7059, USA}
\author{Mikhail~A.~Stephanov}
\affiliation{Department of Physics, University of Illinois, Chicago,  IL 60607-7059, USA}

\date{March 2010}


\begin{abstract}
  We build a ``bottom-up'' holographic model of charmonium by matching
  the essential spectral data. We argue that this data must include
  not only the masses but also the decay constants of the $J/\psi$ and
  $\psi'$ mesons. Relative to the ``soft-wall'' models for {\em light}
  mesons, such a matching requires two new features in the holographic
  potential: an overall upward shift as well as a narrow ``dip'' near
  the holographic boundary. We calculate the spectral function as well
  as the position of the complex singularities (quasinormal frequencies) of
  the retarded correlator of the charm current at finite
  temperatures. We further extend this analysis by showing that the
  residues associated with these singularities are given by the
  boundary derivative of the appropriately normalized quasinormal
  mode. We find that the ``melting'' of the $J/\psi$ spectral peak
  occurs at a temperature $T\approx 540$ MeV, or 2.8 $T_c$, in good
  agreement with lattice results.
\end{abstract}

\maketitle

\section{Introduction}
\label{sec:intro}

The fate of quarkonium states such as $J/\psi$ in the quark-gluon plasma
has been a subject of continuous interest for both theorists and
experimentalists since the famous proposal~\cite{Matsui:1986dk} to use
the suppression of $J/\psi$ production in heavy-ion collisions as a
signature of the quark-gluon plasma formation. The screening of color
forces should weaken or prevent $c\bar c$ pair binding, manifesting in
$J/\psi$ suppression in heavy-ion collisions capable of creating
quark-gluon plasma. Making detailed predictions for this effect is a
formidable challenge, requiring understanding of such factors as
$c\bar c$ pair production, binding, and survival of charmonium states
in medium, as well as their interplay in the dynamically evolving high
energy density medium (see, e.g., Ref.~\cite{Satz:2005hx} for review).

A field-theoretical quantity which encodes much, but certainly not
all, information necessary in this regard is the spectral
function of the operator of charm current, $J^\mu=\bar c\gamma^\mu
c$. This operator, acting on the vacuum state, produces charmonium
$1^{--}$ states such as $J/\psi$, $\psi'$, etc. By studying the
response of the thermal medium to the action of such an operator we
can learn about the fate of the charmonium states in this medium. The
dissociation, or ``melting'', of the charmonium states manifests
itself in the gradual decrease with temperature of the strength of the
response at the frequencies at which the vacuum state would
``resonate'' by creating a charmonium bound state.

The current-current correlator and its spectral function are
field-theoretically
 well-defined quantities, interesting in their own right. They can be, in
principle, calculated using {\it ab initio\/} lattice Monte Carlo
approach.  The task of extracting the charmonium spectral functions
from discrete Euclidean correlators measured on the lattice is
difficult, but significant progress has been made recently using the
Maximum Entropy Method
(MEM)~\cite{Umeda:2000ym,Umeda:2002vr,Asakawa:2003re,Datta:2003ww,Jakovac:2006sf,Aarts:2007pk,Bazavov:2009us,Ding:2009ie}.
The spectral peak corresponding to the $J/\psi$ state appears to be
still prominent up to temperatures as high as almost 2.5 $T_c$, where
$T_c\approx 190$ MeV (see, e.g, Ref.~\cite{Bazavov:2009zn} for a review) 
is the critical/crossover temperature at which
the thermal QCD medium undergoes deconfining transition.  To
understand lattice results better, it is very desirable to have
analytically controllable models which could allow us to study the
``melting'' of the quarkonium spectral peak.

The models based on a non-relativistic Schr\"odinger potential have been
traditionally used for this purpose. In the pioneering work of
Ref.~\cite{Karsch:1987pv}, the effective temperature-dependent potential was
modeled by introducing Debye screening into the linearly confining
potential. The recently studied models have used a non-relativistic
Schr\"odinger potential, whose dependence on temperature is introduced
either by input from the
lattice~\cite{Mocsy:2005qw,Wong:2006bx,Cabrera:2006wh,Alberico:2006vw,Mocsy:2007jz,Mocsy:2007yj,Mocsy:2008eg,Mocsy:2009ca,Petreczky:2010yn},
or resummation-improved perturbation theory~\cite{Laine:2006ns,Laine:2007gj,Burnier:2007qm}.

The non-relativistic thermal potential models typically predict
charmonium dissociation at temperatures approximately 1.2--1.5 $T_c$
\cite{Mocsy:2007yj}, which appear to be too low compared to the
dissociation temperatures inferred from the lattice. Several possible
resolutions of this puzzle have been proposed (as reviewed, e.g., in
Refs.~\cite{Mocsy:2007yj,Young:2008he,Rapp:2009az}).

Gauge-gravity holographic
correspondence~\cite{Maldacena:1997re,Gubser:1998bc,Witten:1998qj}
provides a new field-theoretically consistent framework for modeling
the properties of thermal strongly-coupled medium (see,
e.g.,~\cite{Gubser:2009fc,Gubser:2009md} for recent reviews). Although
{\it ab initio\/} calculations in QCD-proper are still not possible
within this approach, many properties of QCD, both in vacuum and at
finite temperature can be modeled. ``Top-down'' reduction of string
theory on suitably chosen backgrounds allows one to study QCD-like
field theories, such as theories with flavored matter in
fundamental color
representation~\cite{Karch:2002sh,Sakai:2004cn,Sakai:2005yt,Mateos:2006nu}. 
Alternatively,
reversing the rules of the holographic correspondence one can build
``bottom-up'' models by matching the relevant features of QCD, such as
chiral symmetry breaking, operator product expansion constraints on
correlators and linear confinement~\cite{Polchinski:2001tt,Son:2003et,Brodsky:2003px,deTeramond:2005su,Erlich:2005qh,Da
  Rold:2005zs,Karch:2006pv}.

The fate of meson states in strongly coupled thermal medium has been
a subject of many
studies~\cite{Babington:2003vm,Hoyos:2006gb,Mateos:2006nu,Peeters:2006iu,Liu:2006nn,Erdmenger:2007cm,Ejaz:2007hg,Mateos:2007vn,Myers:2008cj,Paredes:2008nf,Evans:2008tv,Erdmenger:2009ce}
using a ``top-down'' approach to implementing massive flavor degrees
of freedom introduced in
Refs.~\cite{Karch:2002sh,Kruczenski:2003uq}. Spectral functions and
quasinormal modes were studied in
Refs.~\cite{Hoyos:2006gb,Peeters:2006iu,Myers:2008cj,Evans:2008tv,Erdmenger:2009ce}. Structure
of quarkonium in these theories
was also studied in~\cite{Hong:2003jm}.

It is reasonable to expect that reproducing the charmonium spectrum
correctly is an important prerequisite for predicting its finite
temperature properties.  One important way in which the spectrum of
heavy quarkonium excitations in ``top-down'' models differs from QCD
is that in these models the lowest meson mass $m_1$ emerges as the
only scale determining the masses of both the ground state and the
excited mesons \cite{Kruczenski:2003be}: $m_n\sim nm_1$, $n\gg 1$.  In
contrast, in the heavy quark limit of QCD, the mass of the ground
state, such as $J/\psi$, is controlled by a parameter (heavy quark
mass) different from the parameter (string tension, or $\Lambda_{\rm
  QCD}$) controlling the level spacing of excited states: $m_n^2\sim
m_1^2+n\Lambda_{\rm QCD}^2$,~Refs.~\cite{pdg,Gershtein:2006ng}. It
would be interesting to find a holographic model which could
represent these features of quarkonium spectrum correctly.

\section{Approach and Outline}
\label{sec:approach}

In this paper, we shall approach this problem from the complementary
side of the ``top-down'' models. Taking the quarkonium properties at
zero temperature as input, we shall construct the holographic dual
which matches these properties in the spirit of the ``bottom-up''
approach of Refs.~\cite{Erlich:2005qh,Da
  Rold:2005zs,Karch:2006pv}. Then we shall study how such a model
would evolve with temperature.

The pioneering calculation of the $J/\psi$ spectral functions within such an
approach has been performed in Ref.~\cite{Fujita:2009wc,Fujita:2009ca}.
Building on the success of the ``bottom-up'' models for {\em light}
mesons~\cite{Erlich:2005qh,Da Rold:2005zs,Karch:2006pv} the authors of
Ref.~\cite{Fujita:2009wc} changed the scale in the ``soft-wall'' model
of Ref.~\cite{Karch:2006pv} to match the mass of $J/\psi$ meson, in
place of the $\rho$ meson.\footnote{Similar ``rescaling'' has been
  used in earlier papers in order to calculate the charmonium
  dissociation temperature, Refs.~\cite{Kim:2007rt,Hou:2007uk}, but
  not the spectral functions.}  However, such an approach suffers from
the drawback already discussed above for the ``top-down'' models: the
scale that sets the level spacing of quarkonium excited states is the
same as the scale that sets the mass of the ground state, in effect
$\Lambda_{\rm QCD}\sim m_{J/\psi}$ in the models of
Refs.~\cite{Fujita:2009wc,Fujita:2009ca,Kim:2007rt,Hou:2007uk}. What we
would like instead is a spectrum which has a gap between the vacuum and
the ground state whose scale is set by a parameter independent from
the spacing of higher excited levels. Experimentally, the slope of the
radial excitation trajectory for $J/\psi$ is similar to that of the
$\rho$ mesons~\cite{pdg,Gershtein:2006ng}, and we want the model to
match that property of QCD.

The matching of the quarkonium mass spectrum is not the only new
ingredient which we need to introduce. We point out that another important
quantity to be matched is the {\em decay constant} of the quarkonium. Simply
put, in order for the spectral function to be correct at high
temperature, we should begin with a function correct at zero
temperature, which requires matching not only the position, but also
the strength of the $J/\psi$ resonance, {\it i.e.} the decay constant. 

One of the most striking
qualitative consequences of such a more realistic model of quarkonium
is a much more robust spectral peak of $J/\psi$, which persists up to
temperature 540 MeV, i.e., 2.8 $T_c$, in agreement with
lattice studies. This is in contrast to Ref.~\cite{Fujita:2009wc}
which finds charmonium peak ``melting'' at around 1.2 $T_c$. We take
better agreement of our results with the lattice calculations as
evidence that the model we introduce captures essential features of
quarkonium.

In Section~\ref{sec:setup}, to make the paper more self-contained, we
lay out the general setup of the holographic model.
%The reader familiar with the basics of holographic QCD can skip Section~\ref{sec:setup} and go
%directly to 
Section~\ref{sec:hologr-potent} discusses the
features of the holographic potential we introduce to model charmonium. 
Section~\ref{sec:finite-T} applies the methods of
holographic QCD at finite temperature to our model and, 
similar to Section~\ref{sec:setup}, should be
familiar to practitioners. The finite temperature spectral functions
obtained using the proposed new holographic potential are presented in
Section~\ref{sec:results}. In Section~\ref{sec:qnm}, we further our
study of the thermal properties of charmonium by considering
quasinormal modes.  To facilitate this analysis, in
Section~\ref{sec:residue}, we generalize the quantitative relation between the
holographic wave-function and the residue at the corresponding
singularity of the
2-point correlator (i.e., decay constant)
to the case of quasinormal modes at nonzero
temperature. Interestingly, this relationship is the same as the known
one in vacuum, 
after the quasinormal mode is appropriately
normalized. We derive the expression for such a norm in
Section~\ref{sec:residue}.  In Section~\ref{sec:analysis}, we use the
quasinormal modes and the residues to analyze the charmonium spectral
function more quantitatively. We conclude in Section~\ref{sec:disc}
with a summary and a discussion of the results.

\section{Setup}
\label{sec:setup}

We shall focus on the two-point correlation function of the heavy quark (charm) current, 
$J^{\mu} =
\bar{c}\gamma^{\mu}c$, and define the retarded Green's
function:
\begin{equation}
  \label{eq:G_R}
  G_R(q) = - i \int\! d^4x\, e^{iqx}\,\theta(x_0)\, 
\left\langle\,[J^\perp(x), J^\perp(0)]\,\right\rangle,
\end{equation}
where $J^\perp$ is a component of the current orthogonal to the 4-vector
$q^\mu$: $q_\mu J^\mu=0$ and $\langle\ldots\rangle$ denotes thermal average.
For simplicity, we shall consider only the case $\bm q=0$, i.e., $q=(\omega,\bm 0)$. Thus
$J^\perp$ will denote any spatial component of $J^\mu$, e.g.,
$J^x$. We shall also define  $G_R(\omega)\equiv G_R(\omega,\bm 0)$. 
%  The function
% $G_R(\omega)\equiv G_R(\omega,\bm 0)$ obeys reflection symmetry
% $G(-\omega^*)=G(\omega)^*$.
% as a function of $\omega^2$, this
% becomes Schwartz reflection symmetry has a branch cut along the real axis of $\omega$.
% The discontinuity across this cut is purely imaginary, odd function
% of $\omega$. The spectral density $\rho(\omega,\bm q)$ as half of
% that discontinuity.
The spectral function is defined, for real values of $\omega$, as
\begin{equation}
  \label{eq:rho-def}
  \rho(\omega) \equiv \rho(\omega,\bm 0) 
= -\im G_R(\omega).
\end{equation}
Among standard properties of $\rho(\omega)$, which can be
shown using spectral representation, is that $\rho(\omega,\bm q)$ equals the Fourier
transform of $\langle [J^\perp(x),J^\perp(0)]\rangle$, and that
$\rho(\omega)/\omega>0$ for all~$\omega$.

In the spirit of the holographic approach, we shall assume that the
generating functional of the heavy quark vector current $J^{\mu}$ can
be represented by the effective action obtained by integrating over a
bulk 5D gauge field $V_M$ (dual to the current) at given fixed
boundary values (equal to the source of the current). The action for
the 5D gauge field is given by
\begin{equation}
  \label{eq:action-5d}
  S = -\frac{1}{4g_5^2}\int d^5x \sqrt{g}\, e^{-\Phi}V_{MN}V^{MN},
\end{equation}
where $g_5^2$ is the 5D gauge coupling and 
$V_{MN} = \partial_M V_N - \partial_N V_M $. The 2-point current correlator is
given by the linear response of the field $V_M$ to an
infinitesimal perturbation of its boundary condition.

We choose the conformally flat representation for the 5D background metric $g_{MN}$
with 4D Lorentz-isometry:
\begin{equation}
  \label{eq:ds2}
  ds^2\equiv g_{MN}\,dx^Mdx^N
=e^{2A(z)}\left[ \eta_{\mu\nu}dx^\mu dx^\nu - dz^2  \right] \ ,
%\quad e^{A}=z^{-1},
\end{equation}
where $\eta_{\mu\nu}={\rm diag}(1,-1,-1,-1)$ is the Minkowski metric
tensor.
The effect of confinement is represented by the non-trivial background
profile of the scalar field $\Phi$ in Eq.~(\ref{eq:action-5d}) in the
same way as it is done in the soft-wall model with dilaton background
in Ref.~\cite{Karch:2006pv}. We shall assume that the physics
associated with the mass of the quark and chiral condensate operator
$\bar c c$ can be also represented as a contribution to the background
field~$\Phi$. Such a contribution is essential if the model is to be
extended to describe different flavors, which must have different
holographic backgrounds, depending on their respective masses. In a
putative top-down approach this contribution could arise through the
interaction of the gauge field with the scalar (``tachyon'') field
dual to the quark mass operator $\bar cc$, e.g., by a mechanism
explored in Refs.~\cite{Erkal:2009xq,Sen:2004nf}.

Following the rules of the holographic correspondence, we shall
calculate the generating functional for correlation functions of the
heavy-quark current by evaluating the action at its extremum for given
boundary conditions. The extremum is given by the solution of the
equations of motion, which in $V_5=0$ gauge read:
\begin{equation} \label{eq:eom-V}
\partial_z[e^{B\left(z\right)} \partial_z V] + q^2 e^{B\left(z\right)}V = 0 \ ,
\end{equation}
where $V$ is any of the three components $V_\perp$ of $V_\mu(q,z)$
transverse to 4-vector $q^\mu$ ($q^\perp=0$) and
\begin{equation}
  \label{eq:B-Phi-A}
   B=A-\Phi\,. 
\end{equation}

Discrete values of $q^2=m_n^2$, for which the Eq.~(\ref{eq:eom-V}) possesses a normalizable solution $V=v_n(z)$ satisfying the
boundary condition $V|_{z=0}=0$, correspond to the masses $m_n$
of the charmonium states, $n=1,2,\ldots = J/\psi,\psi',\ldots\ $. We normalize such solutions as
\begin{equation}
\label{eq:norm}
  \int_0^\infty dz\, e^{B(z)}v_n(z)^2=1\ .
\end{equation}

Back on the field theory side, the decay constants $f_n$ are defined
via matrix elements of the charmonium current between vacuum and a
given vector charmonium state $n$ with polarization $\epsilon_\mu$ as
\begin{equation}
  \label{eq:decay-const-def}
  \langle 0| J_\mu(0)|n\rangle = f_n m_n \epsilon_\mu\ .
\end{equation}
 Holographic correspondence
relates these constants to the (second) derivative of the normal
mode $v_n$ (see Refs.~\cite{Son:2003et,Erlich:2005qh}):
\begin{equation}
  \label{eq:decay-consts}
f_n = \frac1{g_5 m_n}\, v_n'(z)e^{B(z)}\bigg|_{z\to0}\ .
%
%= \frac1{g_5 m_n}\, v_n''(0).
\end{equation}

As one can see from Eqs.~(\ref{eq:eom-V})--(\ref{eq:decay-consts}),
the masses and decay constants of the charmonium states are determined
by the combination $B$ of the gravity ($A$) and dilaton/tachyon ($\Phi$)
backgrounds~(\ref{eq:B-Phi-A}).
In the spirit of the bottom-up approach, we shall choose the function
$B$ so as to satisfy a number of phenomenological and
field-theoretical QCD constraints. We assume that such a background
arises dynamically, but do not attempt to model the corresponding
dynamics. 

We require
that the ultraviolet behavior of the current-current correlator is
conformal, {\it viz.} $G_R(\omega)\sim\omega^2\log(-\omega^2)$ as $\omega^2\to-\infty$,
which translates into:
\begin{equation}
  \label{eq:B-small-z}
  e^{B(z)}\xrightarrow{z\to0} z^{-1}\ ,
%= z^{-1}(1+{\cal O}(z^4)),
%\quad\mbox{as}\quad z\to0,
\end{equation}
and matches that of QCD, which fixes~\cite{Son:2003et,Erlich:2005qh,Da Rold:2005zs}
\begin{equation}
  \label{eq:g_5-Nc}
 g_5^2=12\pi^2/N_c\ .
\end{equation}
% In this paper we shall not be requiring matching of the ${\cal O}(z^4)$ corrections,
% which would correspond to power corrections to the OPE of the
% current-current correlator.

By performing a Liouville transformation, 
\begin{equation}
  \label{eq:Liouville-0}
  \Psi = e^{B(z)/2} V\ ,
\end{equation}
we can bring Eq.~(\ref{eq:eom-V}) to
the canonical Schr\"odinger-like form
\begin{equation}
  \label{eq:Schroedinger-0}
 - d^2\Psi/dz^2 + U(z)\Psi = q^2\Psi\ ,
\end{equation}
with the holographic potential given by
\begin{equation}
  \label{eq:U-B-0}
  U(z) = 
\frac{B''(z)}2 + \left(\frac{B'(z)}{2}\right)^2\ .
\end{equation}

We shall now choose $B$ or, equivalently $U$, to match not only the
mass of $J/\psi$, but also the mass of $\psi'$ as well as (and more
importantly) the decay constants of $J/\psi$ and $\psi'$.

\section{Holographic potential}
\label{sec:hologr-potent}

In order to explain the role and motivation for the features of the
holographic Schr\"odinger potential which we are led to introduce in this
paper, we shall add those features one at a time.

To begin with, it is easy to implement the two-scale excitation
spectrum, correcting the drawback of Ref.~\cite{Fujita:2009wc}, by
adding a constant, $c^2$, to the potential. Taking the potential from
the soft-wall model, $U_{(a)}$, which reproduces the equidistant
mass-squared spectrum controlled by parameter $a$, we thus consider
\begin{equation}
  \label{eq:V-const-shift}
  U_{(a,c)}= U_{(a)}+c^2=
3/(4z^2) + (a^2 z)^2 + c^2\ .
\end{equation}
For this potential, the mass of the ground state is controlled by $c$, independent from
$a$.\footnote{The first term  is the potential for a
conformally invariant theory. It can be, curiously, viewed as a
centrifugal term for a radial Schr\"odinger equation in 2d with the
radial quantum number $m=1$ (coinciding with the spin of the mesons we
study~\cite{Karch:2006pv}).}

The shifted potential in Eq.~(\ref{eq:V-const-shift}), however, has an
important drawback, which is significant as far as the finite
temperature properties of the quarkonium are concerned. Since those
properties are encoded in the spectral function of quarkonium at
finite temperature, it is clear that to get those properties right on
a quantitative level, one must begin with correct spectral function at
{\em zero} temperature. The spectral function at zero temperature
consists of a series of peaks, representing the quarkonium
states. These peaks are characterized not only by their position
(masses) but also by their strengths, i.e., the decay constants of the
quarkonium states. If we take the model of Ref.~\cite{Fujita:2009wc}, we
find that the decay constant of $J/\psi$ is underpredicted by 20\% and
that of $\psi'$ is overpredicted by 15\% -- see
Table~\ref{tab:models}. For the shifted potential in
Eq.~(\ref{eq:V-const-shift}), the situation is much worse: the decay constants
are underpredicted by a factor of order $2-3$.

\begin{table}[ht]
  \begin{center}
    \begin{tabular}{cccc}
      \hline \hline
      &Experiment& $U_{(a)}$ & $U_{(a,c)}$ \\
      Observable& (MeV)& (MeV)& (MeV) \\
      \hline
      $m_{J/\psi}$&3096&3096*&3096*\\
      $m_{\psi'}$&3685&4378&3685*\\
      \hline
      $f_{J/\psi}$&416&348&145\\
      $f_{\psi'}$&296&348&173\\
      \hline \hline
    \end{tabular}
  \end{center}
  \caption{Comparison of the masses and decay
    constants obtained using the scaled soft-wall (as in
    Ref.~\cite{Fujita:2009wc}) and shifted soft-wall
    potentials in Eq.~(\ref{eq:V-const-shift}) with experimental values,
    Ref.~\cite{pdg}. Analytic formulas from Ref.~\cite{Karch:2006pv} are used:
    $m_n^2=4a^2n+c^2$ and $f_n=\sqrt{8n}a^2/(g_5m_n)$, $n=1,2$. The observables
    which are fitted to determine parameter $a$ in $U_{(a)}$ and $a$,
    $c$ in $U_{(a,c)}$ are marked by
    asterisks. Both models significantly underpredict the decay
    constant of $J/\psi$. The shifted potential $U_{(a,c)}$ is worse
    because the parameter $a$ needed for the fit is smaller.}
\label{tab:models}
\end{table}

What should be done to the potential (\ref{eq:V-const-shift}) to
\emph{increase} the value of the $J/\psi$ decay constant? To understand the
solution to this problem, one should recall the relationship between the decay
constant and the rate of change of the holographic wave-function at the
boundary $z\to0$, Eq.~(\ref{eq:decay-consts}):
\begin{equation}
  \label{eq:f-via-psi}
  f_n = \frac1{g_5m_n}\, v_n''(0)
= \frac1{g_5m_n}\, {(\sqrt z\,\psi_n)''}\bigg|_{z\to0}\ ,
\end{equation}
where we used Eq.~(\ref{eq:B-small-z}) and the normalizable
solution $\psi_n$ to Eq.~(\ref{eq:Schroedinger-0}), $\psi_n=e^{B/2}v_n$.
That means the steeper the wave-function $\psi_n$ is near the boundary
$z=0$, the larger is the decay constant. For example, increasing
parameter $a$ would make the function steeper, by squeezing its
support. But that will unacceptably increase the radial trajectory
slope, $dm_n^2/dn$, which we want to preserve. A simple and natural
solution is to provide an ``incentive'' for the wave function to be
larger only in a narrow region close to the boundary, thus forcing it
to approach the boundary value $\psi_n(0)=0$ steeper. This can be
easily achieved by creating a ``dip'' in the potential at small $z\sim
z_d$.

What should be the {\em shape} of the dip at $z_d$? In this paper we shall
consider a minimalistic example of the potential necessary to
match the quarkonium spectral data. We shall use the simplest
approximation for a narrow dip: a negative Dirac delta function.%
\footnote{Replacing the delta-function with a narrow square well
  produces similar results.}

\begin{figure}
  \centering
  \includegraphics[width=.45\textwidth]{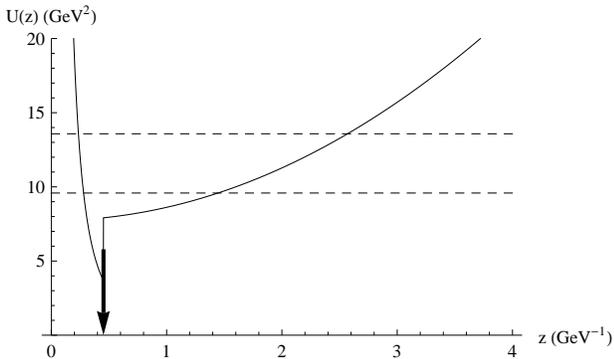}
  \caption{Holographic Schr\"odinger potential at $T=0$ given by
    Eqs.~(\ref{eq:piecewise-U}) and (\ref{eq:parameters}). The negative
    delta-function at $z=z_d$ is indicated by an arrow. Dashed horizontal
    lines show the two lowest
    eigenvalues, $m_1^2$ and $m_2^2$ ($J/\psi$ and $\psi'$). The discontinuity at $z=z_d$
    plays less significant role compared to the delta-function itself.
  }
  \label{fig:pot0}
\end{figure}

Rather than superimposing the delta function onto the potential
$U_{(a,c)}$ in Eq.~(\ref{eq:V-const-shift}), we first observe that, since the
potential undergoes a qualitative change in behavior at $z=z_d$,
manifested in the delta function, there is no reason to expect that it
is described by a function continuous across $z_d$. A more natural, and
convenient, choice is to consider the following piecewise analytic
function:
%\begin{equation}
  \begin{multline}
    \label{eq:piecewise-U}
    U(z) = \frac3{4z^2}\,\theta(z_d-z) + \left((a^2 z)^2 + c^2\right)
    \,\theta(z-z_d) \\- \alpha\delta(z-z_d),
  \end{multline}
%\end{equation}
which represents the necessary large and small $z$ behavior.
There are 4 parameters in this potential, and we use them to fit 4
experimental data points: the masses and the decay constants of
$J/\psi$ and $\psi'$ in Table~\ref{tab:models}. We find (see also Fig.~\ref{fig:pot0})
\begin{equation}\label{eq:parameters}
  \begin{split}
    &a=0.970\mbox{ GeV},\  c=2.781\mbox{ GeV},\\
    &\alpha=1.876\mbox{ GeV},\ z_d=2.211\mbox{ GeV} .
  \end{split}
\end{equation}

\begin{figure}
  \centering
  \includegraphics[width=.45\textwidth]{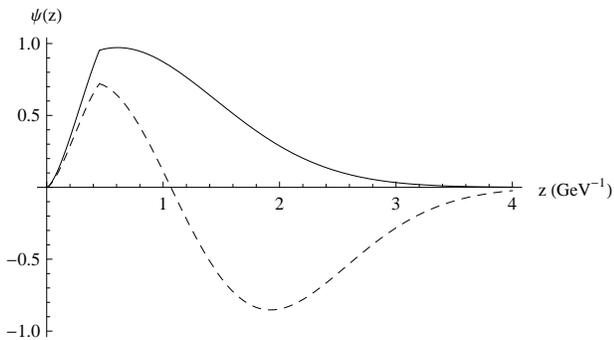}
  \caption{The first two holographic wave functions $\psi_1$ and
    $\psi_2$ (corresponding to $J/\psi$ and $\psi'$ mesons) for the
    potential in Fig.~\ref{fig:pot0}. The kink at $z=z_d$ is due to
    the dip (delta function) in the potential.}
  \label{fig:wave}
\end{figure}
One can see the effect of the delta function on the holographic
wave functions explicitly as shown in Fig.~\ref{fig:wave}. As
expected, the wave functions, varying relatively smoothly for
$z>z_d$, ``dive'' steeply towards their boundary value
$\psi_n(0)=0$ once $z<z_d$.

The potential in Eq.~(\ref{eq:piecewise-U}) can be, certainly,
improved by applying further constraints, e.g., fitting the masses and
decay constants of the higher excited states, and also by applying
constraints which follow from the operator product expansion of the
heavy quark current-current correlator~\cite{Novikov:1977dq} beyond
the leading order.\footnote{In fact, our removal of the $c^2$ term
  from the region of small $z$, achieved by using piecewise potential
  (\ref{eq:piecewise-U}), is motivated, in part, by such
  considerations.}  We shall defer such improvements to future work,
and concentrate here on the consequences of the two major properties
of the potential we introduced: the shift, providing the two-scale
level pattern, and the dip at small $z$, supporting the necessary size
of the decay constants.

From the point of view of the nonrelativistic model of quarkonium, the
effect of increasing the decay constant can be understood as the
effect of decreasing the size of the bound state: the decay constant
is proportional to the probability of finding quark and antiquark at
the same point, which is larger for a more compact state. It is reasonable to
expect that a more compact bound state survives up to higher
temperatures. This is indeed the result we find (see
Section~\ref{sec:results}).

\section{Finite temperature}
\label{sec:finite-T}

Once the zero-temperature parameters of the model are fixed, we wish
to determine the spectral function of the charm current at finite
temperature, as defined on the field theory side by
Eqs.~(\ref{eq:G_R}) and ~(\ref{eq:rho-def}).

On the holographic side, finite temperature correlation functions of
heavy quark current are determined by a similar action
to Eq.~(\ref{eq:action-5d}), but on a different, temperature dependent,
background $\Phi$ and $g_{MN}$, whose main feature is the presence of
a black-hole horizon at some value of $z=z_h$. In a ``top-down''
approach that background would itself be an extremum of the same
gravity/dilaton/tachyon action. We shall not attempt to model the
dynamics of these background fields here, leaving it to further
work. Lacking the equations of motion, which would determine the
background for each temperature, we shall instead make a
minimalistic assumption about the behavior of this background.  

We choose the following representation for the metric consistent 
with the 3d Euclidean spatial isometry,
corresponding to finite temperature:
\begin{equation}
  \label{eq:ds2-BH}
   ds^2=e^{2A(z)}\left[\f dt^2 - d\bm x^2 - \f^{-1} dz^2 \right]\ .
\end{equation}
If the function $\f(z)$ has a simple zero, {\it viz.} $\f(z_h)=0$, the space described by
(\ref{eq:ds2-BH}) possesses an event horizon at $z=z_h$.
The temperature $T$ corresponding to this background is related to $z_h$ as
\begin{equation}
  \label{eq:T-zh}
  T = \frac1{4\pi} \left|\f'(z_h)\right|\ .
\end{equation}
%We shall take a minimalistic ansatz for $\f$, capturing
%the main feature of the finite temperature: the dependence of the
%position of the horizon on $T$ in Eq.~(\ref{eq:T-zh}). 
We assume the
simplest ansatz for $\f$, interpolating between $\f(0)=1$ and $\f(z_h)=0$
with the power $z^4$ dictated by the dimension of the operator of the
energy-momentum:
\begin{equation}
\f=1-(z/z_h)^4\label{eq:f-z4}\ ,
\end{equation}
which turns out to be the same as in the familiar $AdS_5$ black hole solution to
Einstein equations with negative cosmological constant.  The
temperature $T$ corresponding to this background is related to $z_h$
as
\begin{equation}
  \label{eq:zh-T}
  z_h=(\pi T)^{-1}\ .
\end{equation}
As already noted, in principle, we should expect the function
$A(z)$ as well as the background $\Phi(z)$ to depend on the
temperature. Since we do not attempt to model the background
dynamically, we have little choice but to neglect the dependence of
the background $B$ in Eq.~(\ref{eq:B-Phi-A}) on $T$. Our purpose here
is to describe the effect of the temperature semiquantitatively. The
presence of the black-hole horizon in the metric (\ref{eq:ds2-BH}) is
the main property that our ansatz is aimed to capture.

Our approach to introducing temperature dependence here is similar to that of
Ref.~\cite{Fujita:2009wc}.~\footnote{Similar approach was used earlier in
  Refs.~\cite{Ghoroku:2005tf,Ghoroku:2005kg} to study light
  flavor mesons at finite temperature.} In addition to making our model simpler,
this will allow us to see the effect of the features of the potential
we introduce (specifically, the ``dip''), in comparison with
Ref.~\cite{Fujita:2009wc}, more clearly.

Therefore, the equation we solve to determine the correlator of the heavy quark
current at finite temperature is given by
\begin{equation}
  \label{eq:eom-finite-T}
  \partial_z(\f e^{B} \partial_z V) + \omega^2 \f^{-1}e^{B}V = 0 \ ,
\end{equation}
where function $B(z)$ is determined by solving Eq.~(\ref{eq:U-B-0})
 for given $U(z)$.
The retarded correlation function is given, according to the
well-known prescription of 
Ref.~\cite{Son:2002sd,Herzog:2002pc}, by
\begin{equation}
  \label{eq:G_R-V}
  G_R(\omega) = -\frac1{g_5^{2}}\, \f e^B{V'(z,\omega)}\bigg|_{z=\epsilon}=
  -\frac1{g_5^{2}}\,\frac{V'(\epsilon,\omega)}{\epsilon},
\end{equation}
where $\epsilon\to0$ is an ultraviolet regulator and $V(z,\omega)$ is
the solution of Eq.~(\ref{eq:eom-finite-T}) with boundary conditions:
\begin{equation}
  \label{eq:bc-V}
  \begin{split}
&    V(\omega,\epsilon)=1\ ;\quad 
\\ 
&    V(\omega,z)\xrightarrow{z\to z_h}
    C(\omega)(1-z/z_h)^{-i\omega/(4\pi T)}\ ,
  \end{split}
\end{equation}
where, for each $\omega$, $C(\omega)$ is a free constant determined,
if needed, by solving
equation~(\ref{eq:eom-finite-T}). The spectral function can be
calculated as the imaginary part of $G_R(\omega)$ given by
Eq.~(\ref{eq:G_R-V}). Alternatively, one can observe that $\im (\f e^B
V^* \pd_z V)$ is $z$-independent and evaluate it at $z=z_h$, instead
of $z=\epsilon$:
\begin{equation}
  \label{eq:rho-V}
g_5^2\rho(\omega) = \im
{\epsilon}^{-1}{V'(\epsilon,\omega)}
= |C(\omega)|^2 e^{B(z_h)} \omega\ ,
\end{equation}
which demonstrates explicitly that $\rho(\omega)/\omega>0$ condition
is a built-in property of the model.

The Green's function obtained using Eq.~(\ref{eq:G_R-V}) is
logarithmically divergent as $\epsilon\to0$. This logarithmic
dependence on the UV regulator $\epsilon$ matches the expectation on
the field theory side. It does not concern us here because the
imaginary part of Eq.~(\ref{eq:G_R-V}), i.e., the
spectral function given by Eq.~(\ref{eq:rho-V}),
%of the currents defined by Eqs.~(\ref{eq:G_R-V}) and~(\ref{eq:rho-def}) 
is finite, as it should be in the
corresponding field theory.

The solutions of Eq.~(\ref{eq:eom-finite-T}) are easier to study by
applying the Liouville transformation 
\begin{equation}
  \label{eq:Liouville-T}
  \Psi(\z(z)) = e^{B(z)/2}\, V(z);\qquad \z(z) =\int_0^z dz'/\f(z')\ , 
\end{equation}
which yields the equivalent
Schr\"odinger equation for the function $\Psi(\z)$:
\begin{equation}
  \label{eq:Schroedinger-T}
 - d^2\Psi/d\z^2 + U_T(\z)\Psi = \omega^2\Psi\ ,
\end{equation}
where the potential is given by % (cf. Eq.~(\ref{eq:U-B-0}))
\begin{multline}
  \label{eq:U-B-T}
    U_T(\z(z)) =
\frac{d^2 B/d\z^2}2 + \left(\frac{d B/d\z}{2}\right)^2
\\
= \left[ \frac{B''(z)}2 +
      \left(\frac{B'(z)}{2}\right)^2 + \frac{B'(z)\f'(z)}{2\f(z)}
    \right]\f(z)^2\,.
\end{multline}
At $T=0$ this equation becomes Eq.~(\ref{eq:U-B-0}) with $\z=z$. We use
Eq.~(\ref{eq:U-B-0}) with $U(z)$ given by Eq.~(\ref{eq:piecewise-U}) to find
$B(z)$, which then, upon substitution into Eq.~(\ref{eq:U-B-T})
gives us $U_T(\z(z))$.

\section{Results}
\label{sec:results}

\begin{figure}
  \centering
  \includegraphics[width=.45\textwidth]{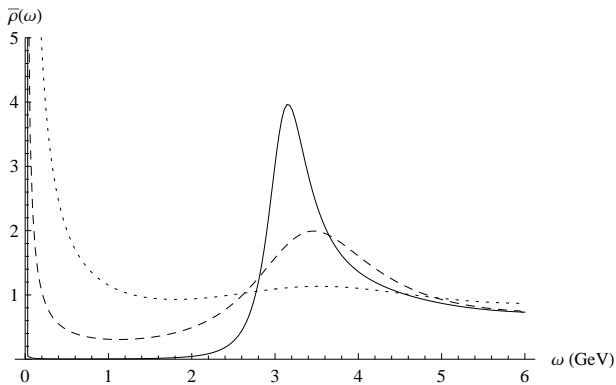}
  \caption{Rescaled spectral function, Eq.~(\ref{eq:rho-bar}), at
    $T=200$ (solid curve), 400 (dashed curve), and 600 MeV (dotted
    curve) in the holographic model of this paper.}
\label{fig:spectral}
\end{figure}

For the purpose of presenting the results, 
it is useful to observe that in a theory with conformal behavior at
short distances, one can expect that
\begin{equation}
\rho(\omega)/\omega^2
\xrightarrow{\omega\to\infty}
\mbox{dimensionless const}.
\label{eq:rho-omega2}
\end{equation}
This large $\omega$
behavior is dominated by the contribution of the
unit operator in the OPE of the product of the currents. 
In QCD, the
dimensionless constant can be calculated, due to the asymptotic freedom,
$
  \rho(\omega)/\omega^2
\xrightarrow{\omega\to\infty}
N_c/(24\pi)
$,
and it is matched in the holographic model by
Eqs.~(\ref{eq:B-small-z}),~(\ref{eq:g_5-Nc}):
\begin{equation}
  \label{eq:rho-asymptotics}
  \frac{\rho(\omega)}{\omega^2}
\xrightarrow{\omega\to\infty}
\frac\pi{2g_5^2}\ .
\end{equation}
Equation~(\ref{eq:rho-asymptotics}) motivates the use of the rescaled
spectral function:
\begin{equation}
  \label{eq:rho-bar}
  \bar\rho(\omega)\equiv 
\frac{\rho/\omega^2}{(\rho/\omega^2)|_{\omega\to\infty}}
=
\frac{2g_5^2}{\pi}\frac{\rho}{\omega^2}\ ,
\end{equation}
such that $\bar\rho(\infty)=1$.
The spectral functions of the charm quark current obtained in our
model at three representative temperatures are shown in
Fig.~\ref{fig:spectral}.

We observe that a pronounced $J/\psi$ spectral peak survives at least
beyond $T=400$ MeV $\approx\ 2\, T_c$. This is in agreement with
lattice results discussed in the introduction.
In comparison, the model in Ref.~\cite{Fujita:2009wc}, predicts the
disappearance of the peak already above $T\sim 1.2\, T_c$. %  Similarly,
% the models based on nonrelativistic Schr\"odinger equations predict
% disappearance of the peak above $T\sim 1.2-1.5\, T_c$.
We can also see that the $J/\psi$ peak has mostly disappeared at $T=600$ MeV. 

Another interesting observation is that the peak at zero $\omega$
grows with temperatures. This peak is related to DC conductivity $\sigma$
 by Kubo formula
 \begin{equation}
\sigma=\lim_{\omega\to0} \rho(\omega)/\omega\ .
\label{eq:kubo}
\end{equation}
It is easy to show, using Eq.~(\ref{eq:rho-V}) and the fact
that $V\equiv 1$ is the solution to Eq.~(\ref{eq:eom-finite-T}) 
at $\omega=0$, that (c.f. Ref.~\cite{Iqbal:2008by})
\begin{equation}
  \label{eq:conductivity}
  \sigma = g_5^{-2} e^{B(z_h)}\ .
\end{equation}
On the field theory side, the growth of the charm conductivity is a
natural consequence of the dissociation of quarkonium bound states.

\begin{figure}
  \centering
  \includegraphics[width=.45\textwidth]{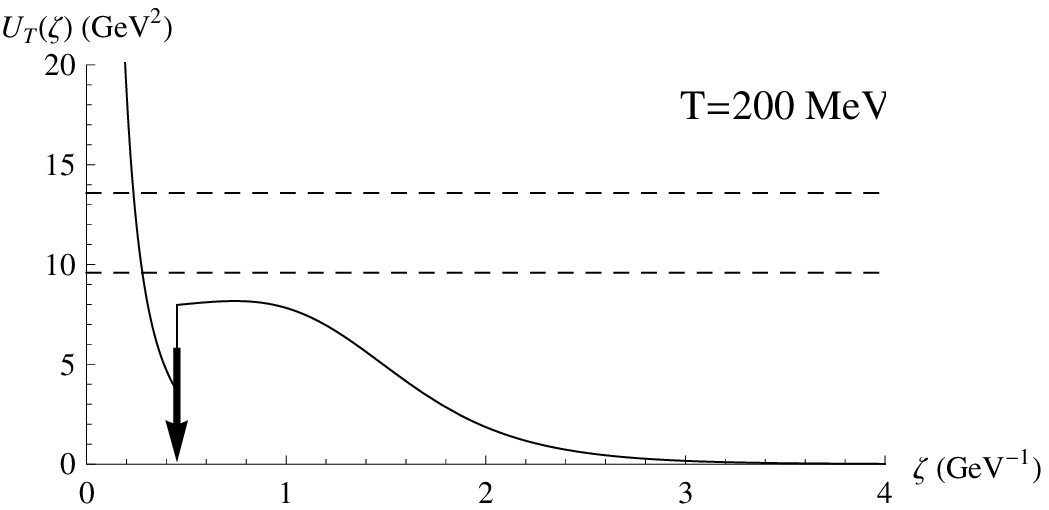}
  % \caption{Holographic potential in Eq.~(\ref{eq:U-B-T}) at $T=200$
  %   MeV. See also Fig.~\ref{fig:pot0}.}
% \label{fig:pot200}
% \end{figure}

% \begin{figure}
%   \centering
  \includegraphics[width=.45\textwidth]{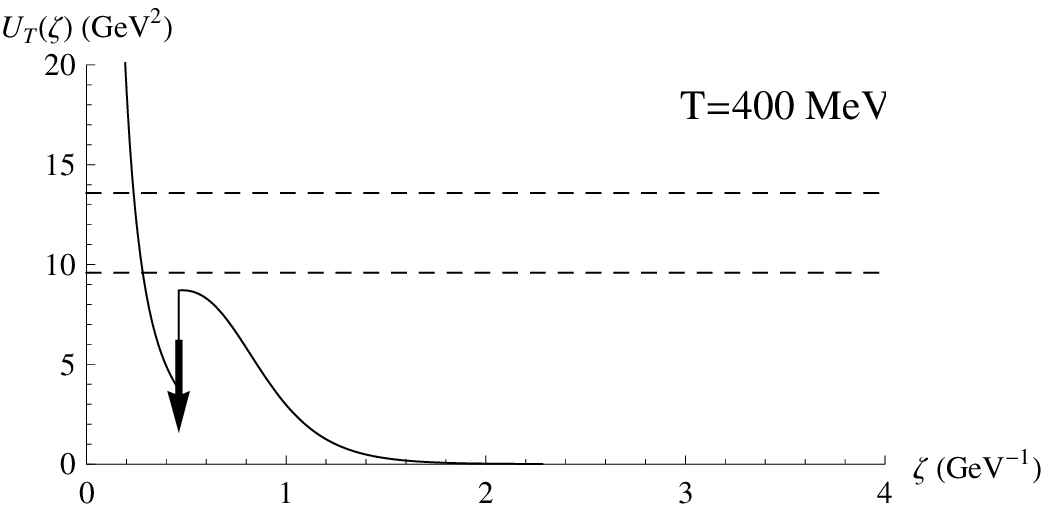}
  \caption{Holographic potential in Eq.~(\ref{eq:U-B-T}) at $T=200$
    and 400
    MeV. See also Fig.~\ref{fig:pot0}.}
\label{fig:pot200-400}
\end{figure}

It is also instructive to look at the temperature evolution of the
holographic potential. Comparing the potentials at $T=200$ and 400 MeV
in Fig.~\ref{fig:pot200-400}, one can see that the
barrier protecting the (quasi-)bound state from decay becomes thinner,
and the depth of the delta function dip becomes smaller with
increasing temperature.

In order to obtain a more objective and quantitative description of
the evolution of the size and shape of the spectral peak, we shall now take
a look at the quasinormal modes defined by
equation~(\ref{eq:eom-finite-T}) with outgoing wave boundary condition
from Eq.~(\ref{eq:bc-V}).

\section{Quasinormal modes}
\label{sec:qnm}

In order to study the properties of the peak in the spectral function,
corresponding to $J/\psi$, we shall use its relation to one of the poles of
the function $G_R(\omega)$ in the complex plane. Such poles correspond
to solutions $V=v_n$ of the equation~(\ref{eq:eom-finite-T}) satisfying
boundary conditions
\begin{equation}
  \label{eq:bc-vn}
  \begin{split}
&    v_n(\epsilon)=0\ ;\quad \\
&    v_n(z)\xrightarrow{z\to z_h}
    c_n(1-z/z_h)^{-i\omega/(4\pi T)}\ ,
  \end{split}
\end{equation}
similar to the b.c. in Eq.~(\ref{eq:bc-V}) at the horizon $z=z_h$, but
not at the boundary $z=\epsilon$. Unlike the b.c. in Eq.~(\ref{eq:bc-V}), the
b.c. in Eq.~(\ref{eq:bc-vn}) are homogeneous and specify the solution
only up to an overall normalization constant.  We describe the
most suitable normalization condition for $v_n$ in
Section~\ref{sec:residue}.

Such solutions $v_n$ are called quasinormal modes due to their
similarity to the normal modes of the $T=0$ equations. Similarly to
normal modes, the quasinormal modes exist only for a discrete set of
frequencies $\omega_n$. However, due to the nature of the b.c. in
Eq.~(\ref{eq:bc-vn}), these frequencies are complex.  

Quasinormal modes in the context of gauge/gravity duality have been
studied extensively, see, e.g.,
Refs.~\cite{Horowitz:1999jd,Starinets:2002br,Nunez:2003eq,Kovtun:2005ev,Myers:2007we},
and in particular, in application to the fate of mesons at finite
temperature in Refs.~\cite{Hoyos:2006gb,Peeters:2006iu,Myers:2008cj,Evans:2008tv,Erdmenger:2009ce}. We
follow a standard numerical method to obtain the values of the
quasinormal frequencies as well as the corresponding eigenfunctions of
the mode equation. In summary, the main challenge arises due to the
fact that, since $\im \omega_n<0$, the boundary condition selects the
solution dominant as $z\to z_h$. Thus, numerical integration of the
equation must be carried out with precision much better than
$(1-z/z_h)^{-\im \omega_n/(2\pi T)}$, to make sure the ``wrong''
solution is not mixed in. This effectively prevents the numerical
integration from reaching $z=z_h$. The standard solution to this
problem \cite{numrec}, and the one we adopt, is to build the solution
in a small but finite interval $[z_h(1-\delta),z_h]$ using a truncated
Taylor (Frobenius) expansion. The numerical integration can then be
carried out starting from the point $z_h(1-\delta)$ away from $z_h$.

\begin{figure}
  \centering
  \includegraphics[width=.45\textwidth]{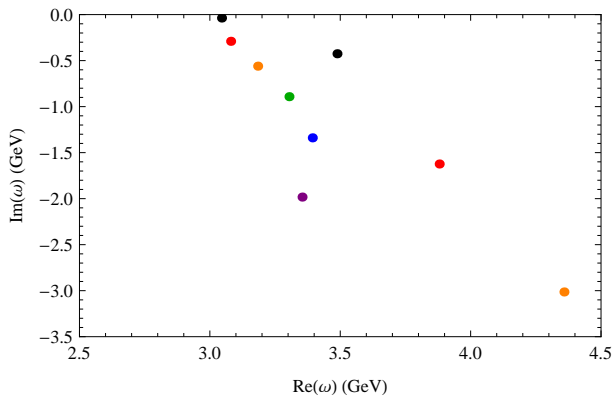}
  \caption{Quasinormal modes $\omega_1$ ($J/\psi$) and $\omega_2$
    ($\psi'$) for a sequence of
    temperatures: $T=100,200,300,400,500,600$ MeV for $\omega_1$ and
    $T=100,200,300$ MeV for $\omega_2$. Colors (online) correspond to
    temperatures.}
  \label{fig:qnm}
\end{figure}

The trajectories of the first two quasinormal frequencies on the
complex plane are shown in Fig.~\ref{fig:qnm}. One can see that the imaginary
part of the second mode $\omega_2$ increases much faster with
temperature than that of $\omega_1$. This corresponds to the
observation that the $\psi'$ peak ``melts'' very early, in accordance with the
sequential dissociation scenario~\cite{Karsch:2005nk}.

We can demonstrate the relationship between the quasinormal
modes and the peak more clearly by calculating the residue $r_n$
of $G_R$ in the pole corresponding to the mode $n$
\begin{equation}
  \label{eq:r_n}
  r_n \equiv \lim_{\omega\to\omega_n}(\omega-\omega_n)G_R(\omega)\ .
\end{equation}
It is easy to see that at $T=0$, the residue is related 
to the $T=0$ decay constant in Eq.~(\ref{eq:decay-consts}) as
\begin{equation}\label{eq:r_n-f_n}
r_n = \frac{f_n^2m_n^2}{2\omega_n}\, 
\qquad (T=0)
\ .
\end{equation}
We shall define the
contribution of the pole to the rescaled spectral function
(\ref{eq:rho-bar}) as
\begin{equation}
  \label{eq:rho_n}
  \bar\rho_n(\omega) \equiv \im \left(\frac{\bar r_n}{\omega-\omega_n}
\right),
\quad\mbox{where}\quad 
\bar r_n=\frac{2g_5^2}{\pi}\,\frac{r_n}{\omega_n^2}\,.
\end{equation}

Due to the well-known property of the Green's function,
$G_R(\omega)^*=G_R(-\omega^*)$, the poles must come in pairs,
$\omega_n$ and $\omega_{-n}\equiv-\omega_n^*$, with residues $r_n$ and
$r_{-n}\equiv -r_n^*$.

In Section~\ref{sec:residue}, we show how the quasinormal mode could in
fact be normalized
and derive a useful expression for the residue $r_n$ in terms of the
corresponding quasinormal mode. The result is an extension
of the $T=0$  equation~(\ref{eq:decay-consts}).

\begin{figure}
  \centering
  \includegraphics[width=.45\textwidth]{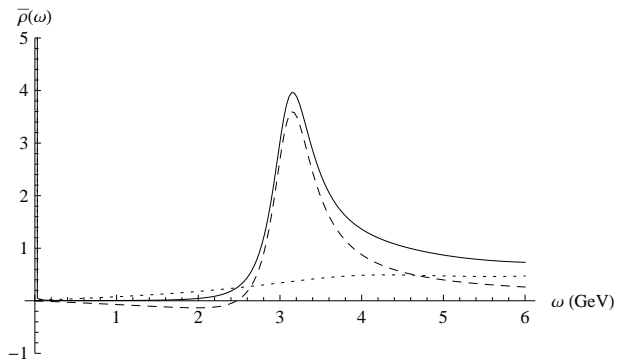}
  \caption{The scaled spectral function at $T=200$ MeV. The dashed line is
    the contribution of the first quasinormal mode,
    $\bar\rho_1+\bar\rho_{-1}$. The dotted line is the result
    of subtraction of that contribution from the spectral function.}
  \label{fig:peak}
\end{figure}

To demonstrate the usefulness of the expression for $r_n$, given by
Eqs.~(\ref{eq:GR-pole}),~(\ref{eq:u-norm}) we derive below, we show in
Fig.~\ref{fig:peak} the spectral function $\bar\rho$, together the
contribution of the pole and its ``mirror,'' $\bar\rho_1+\bar\rho_{-1}$,
as well as the result of the subtraction
$\bar\rho-\bar\rho_1-\bar\rho_{-1}$ which, as expected, leaves only
monotonic background around the location of the peak,
$\omega\approx{\rm Re}\,\omega_1$.~\footnote{ Although, in the case
  we consider, the contribution $\rho_{-1}$ of the ``mirror'' pole at
  $-\omega_1^*$ is almost negligible in the vicinity of the peak at
  $\omega\approx \re \omega_n$, we keep this contribution just to
  preserve the symmetry of the spectral function in
  $\bar\rho_n+\bar\rho_{-n}$.}

\section{Normalization of the modes and the residues}
\label{sec:residue}

In this Section, we describe how the residue $r_n$ for a pole at
quasinormal frequency $\omega_n$ can be calculated given only the
corresponding quasinormal mode $v_n(z)$. We shall show that the
expression is similar to the zero temperature result,
Eq.~(\ref{eq:decay-consts}), for the decay constants $f_n$.  In order
to do that, we would need to be able to generalize the normalization
condition Eq.~(\ref{eq:norm}) at $T=0$ to the case of  quasinormal modes at
finite $T$. We shall indeed obtain this generalization as a byproduct of
our derivation of $r_n$.

Similar to the derivation of the $T=0$ result in
Ref.~\cite{Erlich:2005qh}, we shall introduce the Green's function
$G(\omega;z,z')$ for the mode equation~(\ref{eq:eom-finite-T}):
\begin{equation}
  \label{eq:Green}
    \pd_z\left(\f e^B\pd_zG\right)+\omega^2\f^{-1}e^BG=\delta(z-z')\ ,
\end{equation}
with (homogeneous) boundary conditions similar to Eq.~(\ref{eq:bc-vn})
\begin{equation} \label{eq:bc-G}
  \begin{split}
    &G(\omega;\epsilon,z')=0\ ;\qquad\\
 &G(\omega;z,z')\xrightarrow{z\to
      z_h} C(\omega,z')(1-z/z_h)^{-i\omega/(4\pi T)} \, ,
  \end{split}
\end{equation}
where, $C(\omega,z')$ is an arbitrary function of $\omega$ and
$z'$. It can be easily checked that
\begin{equation}
  \label{eq:V=G}
  V(\omega,z') = -z^{-1} \pd_z G(\omega;z,z')|_{z=\epsilon} 
\end{equation}
satisfies defining equations~(\ref{eq:eom-finite-T})
and~(\ref{eq:bc-V}). 

The Green's function $G$ can be constructed as
\begin{equation}
  \label{eq:Guv}
  G(\omega;z,z') = \frac{u(\omega,z_<)\,v(\omega,z_>)}{\f e^B[u,v]}\ ,
\end{equation}
where $z_{\{<,>\}}=\{\min,\max\}(z,z')$;  $u$ and $v$ are two solutions of
equation~(\ref{eq:eom-finite-T}) each obeying separately one of the
two boundary conditions in Eq.~(\ref{eq:bc-G}):
\begin{subequations}\label{bc-uv}
  \begin{align}
  &u(\omega,\epsilon)=0\label{eq:bc-u}\ ;\\
  &v(\omega,z)\sim (1-z/z_h)^{-i\omega/(4\pi T)}\label{eq:bc-v}
  \mbox{ for } z\to z_h\ .
\end{align}
\end{subequations}
Both boundary conditions (\ref{bc-uv}) are homogeneous and thus
define $u$ and $v$ only up to overall normalization
constants. However, the normalization of either $u$ or $v$ is canceled
out from Eq.~(\ref{eq:Guv}) due to the Wronskian
\begin{equation}\label{eq:W-def}
[u,v]\equiv u\,\pd_zv-v\,\pd_zu\ ,
\end{equation}
in the denominator. Note also that the denominator in
Eq.~(\ref{eq:Guv}) is independent of $z$:  $\pd_z(\f e^B[u,v])=0$. 

The zeros of the Wronskian $[u,v]$ as a function of $\omega$ 
are the poles of the Green's function (\ref{eq:Guv}). The
corresponding complex value of $\omega$ is the quasinormal frequency, $\omega_n$.
The residue associated with such a pole is given by
\begin{equation}
  \label{eq:residue}
\lim_{\omega\to\omega_n}(\omega-\omega_n)G(\omega;z,z') 
=
\frac{u(\omega,z_<)\,v(\omega,z_>)}{\f e^B\pd_\omega [u,v]}\bigg|_{\omega=\omega_n}.
\end{equation}

Since $\f e^B\pd_\omega [u,v]$ is independent of $z$, we can calculate it by
setting $z$ to a convenient value, which happens to be
$z=\epsilon$. At that point $u(\omega,\epsilon)=0$ and $\pd_\omega u(\omega,\epsilon)=0$.
Furthermore, since at $\omega=\omega_n$ the function $v$ is a constant
multiple of $u$, also $v(\omega_n,\epsilon)=0$. This means,
\begin{equation}
  \label{eq:Wdot-Wvdot}
  \f e^B \pd_\omega [u,v] = \f e^B [u,\pd_\omega v]\bigg|_{z=\epsilon}
\qquad(\mbox{at } \omega=\omega_n).
\end{equation}

Differentiating Eq.~(\ref{eq:eom-finite-T}) obeyed by $v$ with respect to
$\omega$, we obtain the equation for $\pd_\omega v$
\begin{equation}
  \label{eq:dotv}
  \pd_z\left(\f e^B\pd_z(\pd_\omega v)\right)+\omega^2\f^{-1}e^B(\pd_\omega v)+2\omega \f^{-1}e^Bv=0.
\end{equation}
We multiply Eq.~(\ref{eq:dotv}) by $u$, then take Eq.~(\ref{eq:eom-finite-T})
obeyed by $u$ and multiply it by
$\pd_\omega v$, and then subtract the two resulting equations to find
\begin{equation}
  \label{eq:uv=W'}
 \pd_z(\f e^B [u,\pd_\omega v]) = - 2\omega \f^{-1}e^B uv\ .
\end{equation}

Integrating Eq.~(\ref{eq:uv=W'}) over $z$ from $z=\epsilon$ to
$z=z_h(1-\delta)$ and using Eq.~(\ref{eq:Wdot-Wvdot}) we find (at $\omega=\omega_n$)
 \begin{multline}\label{eq:dW/domega=int}
  \f e^B\pd_\omega [u,v] 
= 2\omega\int_\epsilon^{z_h(1-\delta)}\!\!\frac{dz}{\f}\,e^Buv
\\
+ \f e^B [u,\pd_\omega v]\bigg|_{z=z_h(1-\delta)}\,.
\end{multline}
As $\delta\to0$, we can (provided $-\im\omega_n<2\pi T$,
see below) replace $v$ in the last term by its
asymptotics~(\ref{eq:bc-v}). Thus
\begin{equation}
\label{eq:d-omega-v-asymptotics}
\pd_\omega v\xrightarrow{z\to z_h} -\frac{i}{4\pi
T}\log(1-z/z_h)\,v\,,
\end{equation}
and therefore
\begin{equation}
  \label{eq:Wdelta}
  [u,\pd_\omega v]\bigg|_{z=(1-\delta)z_h}
\xrightarrow{\delta\to0} -\frac{i\log\delta}{4}[u,v] +
\frac{i}{4\delta}uv\ .
\end{equation}

At $\omega=\omega_n$, $u\sim v$ and $[u,v]=0$. Thus we can write, for $\delta\to0$,
Eq.~(\ref{eq:dW/domega=int}) as
\begin{equation}
  \label{eq:dW/domega=int-uv}
   \f e^B\pd_\omega [u,v] 
= 2\omega\int_\epsilon^{(1-\delta)z_h}\!\!\frac{dz}{\f}\,e^Buv
+ i e^B uv\big|_{z=(1-\delta)z_h}.
\end{equation}
If we choose the normalization of $u$ and $v$ at $\omega=\omega_n$
in such a way that $u=v=v_n$ and the r.h.s. of
Eq.~(\ref{eq:dW/domega=int-uv}) equals $2\omega_n$, we find from
Eqs.~(\ref{eq:residue}),~(\ref{eq:V=G}),~(\ref{eq:r_n}) and~(\ref{eq:G_R-V}):
\begin{equation}
  \label{eq:GR-pole}
r_n = \frac1{g_5^2}\frac{(v_n'(\epsilon)/\epsilon)^2}{2\omega_n}\ ,
%   G^R = -g_5^{-2}
% \frac{(v_n'(\epsilon)/\epsilon)^2}{2\omega_n(\omega-\omega_n)}
% + \mbox{ regular terms }.
\end{equation}
where $\omega_n$ is the quasinormal frequency and $v_n$ is the
corresponding quasinormal mode normalized as
\begin{equation}
  \label{eq:u-norm}
  \lim_{\delta\to0}\left[\int_\epsilon^{(1-\delta)z_h}\frac{dz}{\f}e^Bv_n^2
+ \frac{i}{2\omega_n} e^B v_n^2\big|_{z=(1-\delta)z_h}\right]=1\ .
\end{equation}

Note that, at $T=0$, the normalization condition in
Eq.~(\ref{eq:u-norm}) reduces to the normalization in Eq.~(\ref{eq:norm})
for the normal mode. The expression for the
residue in Eq.~(\ref{eq:GR-pole}) also reduces to (\ref{eq:decay-consts}) via
Eq.~(\ref{eq:r_n-f_n}).

It is important to realize, however, that $v_n^2$ is involved in
Eq.~(\ref{eq:u-norm}), not $|v_n^2|$, and that both terms in
Eq.~(\ref{eq:u-norm}) are complex. In particular, that means that
complex condition in Eq.~(\ref{eq:u-norm}) fixes both the magnitude
and the phase of $v_n$.

One could also note that $\int(dz/\f)e^Bv^2=\int d\z\, \psi^2$, where
$\z$ and $\psi$ are the Schr\"odinger variables defined in
Eq.~(\ref{eq:Liouville-T}). Written in terms of $\z$ and $\psi$, the
normalization condition in Eq.~(\ref{eq:norm}) is equivalent to the one
introduced by Zeldovich in the study of the quasi-discrete levels in
quantum-mechanical decays~\cite{z-book}, and used in the theory of
quasinormal modes of cavities~\cite{qnm-review}.

Equation (\ref{eq:u-norm}) has a drawback, which becomes
important when $-\im\omega_n/T$ is sufficiently large. Since $\im\omega_n<0$,
 $v_n\sim(1-z/z_h)^{-i\omega/(4\pi T)}\to\infty$ as $z\to z_h$.
The last term in Eq.~(\ref{eq:u-norm}) serves to cancel the divergence in the
integral. However, the remainder is finite only if $-\im\omega_n<2\pi T$. This
follows from the Frobenius expansion of the solution $v$ near $z=z_h$:
 $v(\omega,(1-\delta)z_h)\sim\delta^{-i\omega/(4\pi T)}(1+O(\delta))$. Substituting
 into Eq.~(\ref{eq:u-norm}), we find that after cancellation of the
 divergent terms $\delta^{-i\omega/(2\pi T)}$, the remaining terms are of
 order $\delta^{-i\omega/(2\pi T)+1}$ and vanish as $\delta\to0$  for
 $-\im\omega_n<2\pi T$. However, for $-\im\omega_n>2\pi T$, the remaining terms are still
 divergent as $\delta\to0$. This drawback can be cured by returning to
 Eq.~(\ref{eq:dW/domega=int}) and keeping more terms in the
 asymptotics of $\pd_\omega v$ in
 Eq.~(\ref{eq:d-omega-v-asymptotics}). In practice, we do  use
 Eq.~(\ref{eq:dW/domega=int}), and calculate the last term on the
 r.h.s. using the truncated Frobenius expansion (which we have to generate
 anyway in order to obtain the solution of the mode equation near
 $z_h$, as explained in Section~\ref{sec:qnm}).

\section{Results of quasinormal mode analysis}
\label{sec:analysis}

The analysis of quasinormal modes allows us to study the fate of the
$J/\psi$ more quantitatively. Rather than fitting the spectral
function $\rho$, as done in Ref.~\cite{Fujita:2009wc}, we can read off the peak's
height and width directly from the values of the residue and the
imaginary part of the quasinormal frequency obtained by solving
Eq.~(\ref{eq:eom-finite-T}), with b.c.~(\ref{eq:bc-vn}) and using
Eq.~(\ref{eq:GR-pole}) and~(\ref{eq:u-norm}).
We {\em define} the peak height $H_n$ as
\begin{equation}\label{eq:H-rho_n}
  H_n\equiv \bar\rho_n(\omega)\bigg|_{\omega=\re\omega_n}
=\frac{\re\bar r_n}{\im\omega_n}\ ,
\end{equation}
where $\bar\rho_n(\omega)$ and $\bar r_n$ are given by Eq.~(\ref{eq:rho_n}).
The width of the peak can be defined, as usual, as
\begin{equation}
\Gamma_n\equiv-\im\omega_n\ .\label{eq:Gamma-Imomega}
\end{equation}
% The area under the peak is then given by
% \begin{equation}
%   \label{eq:area-peak}
%   A_n\equiv\int_{-\infty}^{\infty}\frac{\rho_n(\omega)}{\omega^2}
% = \pi Z_n \Gamma_n = \pi\re \bar r_n\,. 
% \end{equation}

It is instructive to compare our results with those obtained by
applying the same quasinormal mode analysis to the scaled soft-wall
model~\cite{Fujita:2009wc}.% (see potential in Fig.~\ref{fig:pot0}).

\begin{figure}
  \centering
  \includegraphics[width=.45\textwidth]{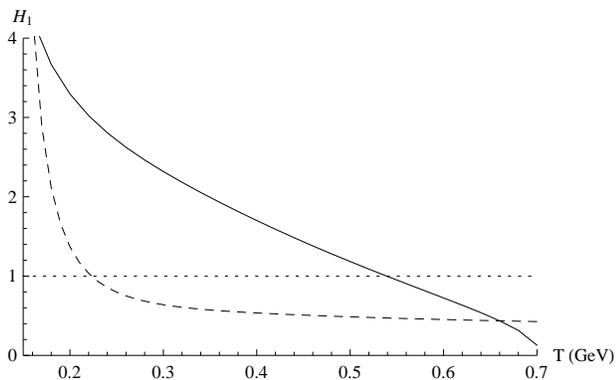}
  \caption{Temperature dependence of the normalized height of the $J/\psi$ 
peak $H_1$ defined in Eq.~(\ref{eq:H-rho_n}) in the holographic
model. For comparison, the same result using the potential
of Ref.~\cite{Fujita:2009wc}, i.e., $U_{(a)}$ of
Eq.~(\ref{eq:V-const-shift}), is shown as a dashed line. The dotted horizontal line shows asymptotic value of~$\bar\rho_n$.}
  \label{fig:height}
\end{figure}

The dependence of the $J/\psi$ peak height $H_1$ on $T$ is shown in
Fig.~\ref{fig:height}. Using $H_1$, one can define a useful criterion
for identifying the ``melting'' temperature, which quantifies
reasonably well the qualitative perception. We can define the
disappearance of the peak as a point at which the height of the peak as
defined by Eq.~(\ref{eq:H-rho_n}) becomes smaller than the asymptotic value
of the background, i.e., $\bar\rho(\infty)=1$.

According to this criterion, the peak survives until temperatures
about 540 MeV, i.e., about 2.8 $T_c$. In contrast, in the rescaled
soft-wall model of Ref.~\cite{Fujita:2009wc} the peak height drops quickly
and disappears at about 230 MeV, or 1.2 $T_c$. The peak is more robust
in our model because of the delta function in the holographic
potential $U$ in Eq.~(\ref{eq:piecewise-U}), which can additionally be checked by reducing $\alpha$ to 0.

\begin{figure}
  \centering
  \includegraphics[width=.45\textwidth]{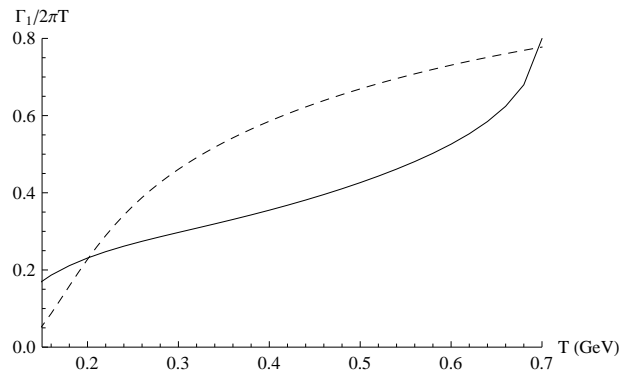}
  \caption{Solid line: the width of the $J/\psi$ peak in units of $2\pi T$ as defined by Eq.~(\ref{eq:Gamma-Imomega}). For comparison, the dashed line shows the
    result using the potential from Ref.\cite{Fujita:2009wc}, i.e.,
    $U_{(a)}$ in Eq.~(\ref{eq:V-const-shift}).}
  \label{fig:width}
\end{figure}

The dependence of the $J/\psi$ peak width $\Gamma_1$ (in units of 
$2\pi T)$ is shown in Fig.~\ref{fig:width}.
Interestingly, the peak begins to broaden somewhat earlier in our
model ($T\sim$100 MeV vs $T\sim$150 MeV). This can be understood as the consequence
of the fact that our potential has a ``softer wall'' at large $z$, and
thus is sensitive to lower temperatures. However, even after the soft wall
is ``melted'' at about 200 MeV, the dip at small $z_d$ continues to
support the $J/\psi$ peak. On the contrary, in the rescaled soft model
Ref.~\cite{Fujita:2009wc}, once the wall has ``melted'', no feature
remains in the potential to support the narrow quasi-discrete $J/\psi$
state.

It is also interesting to note that for $T\to\infty$ the width $\Gamma_1/(2\pi T)\to1$,
which corresponds to the value of the $n=1$ quasinormal frequency,
$\omega_n=2\pi T n(\pm1-i)$ (see, e.g., Ref.~\cite{Myers:2007we}) in the AdS
black-hole background. This agrees with the expectation that at
asymptotically large temperatures the theory should appear almost conformal.

\section{Summary, Discussion and Outlook}
\label{sec:disc}

In this paper, we attempted to build a holographic model of charmonium
using the ``bottom-up'' approach. As was done in
Ref.~\cite{Erlich:2005qh}, we required the matching of the UV behavior
of the charm current correlator to that of QCD, which is known due to asymptotic freedom.
This fixed the small $z$ behavior of the holographic
potential. Similarly, the large $z$ behavior was fixed using the
condition that the meson spectrum is asymptotically equidistant in
mass squared, as in the soft wall model of Ref.~\cite{Karch:2006pv}. 

We then observed that, in QCD, the asymptotic spacing of levels is controlled
by a parameter, $\Lambda_{\rm QCD}$, theoretically
independent from the mass of the ground state, $m_{J/\psi}$. We
incorporated this feature by a shift in the soft wall model
holographic potential. At this
point our model became different from that of
Ref.~\cite{Fujita:2009wc}, where both scales were controlled by one parameter.

We then required that not only the spectrum of the meson masses is matched
to QCD phenomenology, but also the values of the decay constants. This
requirement led us to introduce a ``dip'' in the holographic
potential. We modeled this feature minimalistically, by a delta function.

\begin{figure}[h]
  \includegraphics[width=0.45\textwidth]{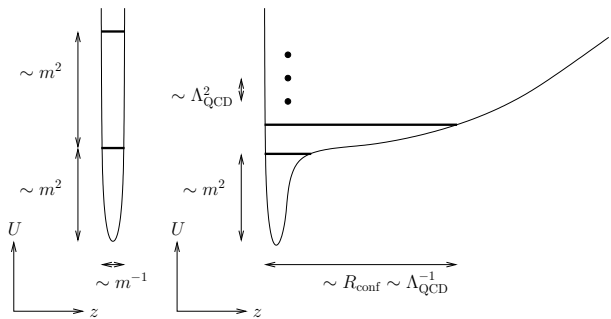}
  \caption[]{A sketch emphasizing the features of the holographic
    Schr\"odinger potentials. {\em Left} -- the ``top-down'' model
    potential (see, e.g., Eq. (3.2) in Ref.~\cite{Paredes:2008nf}) has
    one scale, $m$, characterizing both the ground state mass and the
    level spacing.  {\em Right} -- a potential with separate scales
    for the mass gap, $m$, and for the level spacing, $\Lambda_{\rm
      QCD}$. Due to slower rising wall of the potential at large $z$,
    the holographic wave function is allowed to ``leak'' beyond the
    scale of $m^{-1}$. In this paper the ``dip'' at small $z_d\sim
    m^{-1}$ is approximated by a negative delta function.}
  \label{fig:potentials-sketch}
\end{figure}

A possible way one could anticipate such a feature from a putative
``top-down'' construction is by observing that the existing models
describing heavy mesons do indeed have potentials with the
characteristic scale of the width of the well (flavor brane extension
in the $z$ coordinate) of order of the inverse meson mass $z_d\sim
m^{-1}$.
%  -- small compared to the confinement
% scale, such as confinement radius $R_{\rm conf}\sim\Lambda_{\rm QCD}^{-1}$.
From that point of view, the potential we need
differs from these models by a slower rising ``soft wall'' at large $z$,
permitting the wave-function to ``leak'' significantly beyond scales
of order inverse meson mass $m^{-1}$, and reaching up to 
% scales set by
% the parameter $a^{-1}$, i.e.,
the confinement scale, such as
confinement radius, $R_{\rm conf}\sim\Lambda_{\rm
  QCD}^{-1}$, as illustrated in Fig.~\ref{fig:potentials-sketch}.

It is easy to see that such a modification of the potential is what
accounts for the broadening of the $J/\psi$ peak {\em below} the
dissociation temperature in our model. In contrast, the ``top-down''
models would predict no broadening below dissociation temperature,
though this can change if instanton-like tunneling effects are
considered~\cite{Faulkner:2008qk}.

Having thus constructed a potential with 4 free parameters, we fit those
parameters to the 2 masses and 2 decay constants of $J/\psi$ and $\psi'$.

Obviously, the ``ascetic'' potential which we introduced can be further
improved by matching more than just the 2 states in the charmonium
spectrum. The problem of reconstructing the potential from the spectral
data is a standard inverse problem. The fact that the spectrum
together with the derivatives of the eigenfunctions (initial
velocities, or normalizing constants)
uniquely specifies the potential for a Dirichlet
problem~\cite{inverse-problem-book}, suggests that matching decay
constants is a natural part of such ``bottom-up'' approach.

An interesting related observation is that the same spectral
data can be obtained, in principle, from a non-relativistic model of
quarkonium. It would be interesting to know if a more direct
relationship exists between the holographic potential and the
potential in the non-relativistic Schr\"odinger description of
quarkonia in the heavy quark limit.

It might be also interesting to consider the constraints on the
potential arising from the operator product expansion of the current
correlator~\cite{Novikov:1977dq}. By quark-hadron duality these
constraints can substitute the spectral data. It would be
interesting to explore the finite temperature consequences of these
constraints and compare to the studies of
Ref.~\cite{Morita:2007pt,Morita:2007hv,Morita:2009qk}.

Having specified the model at zero temperature, we consider the
evolution of the spectral function of the charm current with
temperature. Here too, we make a minimalistic assumption about the
temperature dependence of the background, similar to the previous
study of Ref.~\cite{Fujita:2009wc}. The necessity for the assumption
comes from the fact that we do not model the dynamics of the
background itself. This can be remedied in a more comprehensive model
which we leave to further work. We believe that the minimalistic
assumption about $z$-dependence of the temperature factor $h$ and
the scale factor $e^B$ is sufficient to make semiquantitative
predictions at the level of precision generally expected from such a
model approach.

One of the new tools we introduce to study properties of the finite
temperature Green's function is based on the relationship between the
residue in the pole and the holographic wave function, which we
derive, Eqs.~(\ref{eq:GR-pole}),~(\ref{eq:u-norm}). This relationship
is a nontrivial generalization of the zero temperature expression for
the decay constants~\cite{Erlich:2005qh}. This allows us to study not
only the {\em position\/} of the pole in the complex plane, corresponding to
the quasinormal mode, but also the {\em strength\/} of the contribution of
this pole to the spectral function. In particular, this allows us to
devise a simple quantitative criterion for the disappearance of the
peak based on its height relative to the asymptotic level of the
background. Since the dissociation of quarkonium and disappearance of
the peak is not a sharp transition, a criterion of this type might
find its use in comparing results of different approaches in a more
uniform and objective way.

Finally, we find that the new features of the holographic potential, such
as the ``dip'', which we
introduce to model quarkonium more realistically, 
strengthen the robustness of the $J/\psi$
peak, allowing it to persist out to temperatures of about 540 MeV,
i.e., 2.8 $T_c$, according to our criterion.  This is in good
agreement with lattice studies, but not with most non-relativistic
Schr\"odinger models of the quarkonium or the holographic model of
Ref.~\cite{Fujita:2009wc}, which typically predict dissociation at
about 1.2 $T_c$.  The comparison with the lattice results can be made
even more direct by calculating the Euclidean correlators of the
current in the holographic model. We defer this to future work.

\acknowledgments

We would like to thank A.~Karch and D.~Son for discussions. The work of H.G
is supported by DOE ONP contract DE-AC02-06CH11357. The work of
P.M.H. and M.A.S. is
supported by the DOE grant No.\ DE-FG0201ER41195.

%%%%%%%%%%%%%%%%%%%%%%%%%%%%%%%%%%%%%%%%%%%%%%%%%%%%%%%%%%%%%%%%%%%%%%%%%%
%%%%%%%%%%%%%%%%%%%%%%%%%%%%%%%%%%%%%%%%%%%%%%%%%%%%%%%%%%%%%%%%%%%%%%%%%%

\end{document}